# Adversarial Text Generation with Dynamic Contextual Perturbation

Hetvi Waghela
Dep of Data Science
Praxis Business School
Kolkata, INDIA
waghelahetvi7@gmail.com

Jaydip Sen
Dept of Data Science
Praxis Buiness School
Kolkata, INDIA
jaydip.sen@acm.org

Sneha Rakshit
Dept of Data Science
Praxis Business School
Kolkata, INDIA
srakshit149@gmail.com

Subhasis Dasgupta
Dept of Data Science
Praxis Busines School
Kolkata, INDIA
subhasis@praxis.ac.in

*Abstract—* Adversarial attacks on Natural Language Processing (NLP) models expose vulnerabilities by introducing subtle perturbations to input text, often leading to misclassification while maintaining human readability. Existing methods typically focus on word-level or local text segment alterations, overlooking the broader context, which results in detectable or semantically inconsistent perturbations. We propose a novel adversarial text attack scheme named Dynamic Contextual Perturbation (DCP). DCP dynamically generates context-aware perturbations across sentences, paragraphs, and documents, ensuring semantic fidelity and fluency. Leveraging the capabilities of pre-trained language models, DCP iteratively refines perturbations through an adversarial objective function that balances the dual objectives of inducing model misclassification and preserving the naturalness of the text. This comprehensive approach allows DCP to produce more sophisticated and effective adversarial examples that better mimic natural language patterns. Our experimental results, conducted on various NLP models and datasets, demonstrate the efficacy of DCP in challenging the robustness of state-of-the-art NLP systems. By integrating dynamic contextual analysis, DCP significantly enhances the subtlety and impact of adversarial attacks. This study highlights the critical role of context in adversarial attacks and lays the groundwork for creating more robust NLP systems capable of withstanding sophisticated adversarial strategies.

*Keywords—Adversarial Attacks, Natural Language Processing, Contextual Perturbation, Robustness, Pre-trained Language Models, Semantic Fidelity, Misclassification, Text Generation.*

## I. Introduction

The rapid advancement of Natural Language Processing (NLP) has enabled the development of sophisticated models capable of understanding and generating human language with remarkable accuracy. These models, powered by deep learning techniques and vast amounts of data, are now integral to various applications, including sentiment analysis, machine translation, and conversational agents. However, the increasing reliance on these models also raises significant concerns regarding their robustness and security. Adversarial attacks, which involve deliberately crafting input data to deceive machine learning models, have emerged as a critical area of study in ensuring the reliability of NLP systems [1-2].

Adversarial text attacks target NLP models by introducing subtle perturbations to input text, aiming to mislead the model while keeping the changes imperceptible to human readers [3]. These attacks exploit vulnerabilities in the model's understanding and processing of language, revealing potential weaknesses that could be exploited in malicious scenarios. Traditional adversarial attack techniques in NLP often focus on word-level perturbations, such as synonym replacement or character-level alterations. While these methods can be effective, they frequently fail to consider the broader contextual coherence of the text, resulting in perturbations that are either easily detectable or disrupt the overall meaning.

In response to these limitations, we propose a novel adversarial text attack scheme named Dynamic Contextual Perturbation (DCP). The DCP scheme is designed to generate perturbations that are dynamically informed by the contextual environment of the target text. This approach ensures that the adversarial examples not only deceive the NLP model but also maintain semantic fidelity and fluency, making them challenging to detect through traditional defense mechanisms. By leveraging the power of pre-trained language models, DCP intelligently modifies text at various levels, including words, phrases, and sentences, to produce coherent and contextually appropriate adversarial inputs.

*Contributions:* The key contributions of the current work are as follows. First, the proposed scheme DCP has contextual sensitivity unlike many existing methods for adversarial text generation that focus solely on word-level perturbation without considering the context of the text. This enables DCP to dynamically generate perturbations based on the context of the target text allowing more nuanced and contextually relevant alterations. Second, DCP incorporates techniques like synonym replacement, homophone substitution, and paraphrasing to subtly alter the text's meaning while preserving its core semantic content, unlike traditional text attacks that result in perturbed text that lacks semantic coherence. Third, unlike adversarial text generated by existing methods that suffer from poor readability and fluency, DCP produces perturbed text that is indistinguishable from natural language, ensuring that it remains readable and fluent. Finally, DCP employs an adversarial objective function that balances misclassification likelihood with text fluency. By optimizing this objective function through iterative refinement using optimization algorithms, DCP ensures that the generated adversarial samples are effective in fooling NLP models while remaining human-like.

This work aims to contribute to the ongoing efforts in enhancing the security and robustness of NLP systems. By exploring the intersection of context and adversarial perturbations, the DCP scheme not only offers a novel perspective on adversarial text attacks but also sets a foundation for developing more resilient models capable of withstanding sophisticated adversarial strategies.

The structure of this paper is as follows: In Section II, we review related work in the field of adversarial text attacks, highlighting the strengths and limitations of current approaches. Section III details the methodology of the DCP scheme, including the theoretical underpinnings and the algorithmic steps involved in generating adversarial examples. In Section IV, we describe the implementation details, providing insights into the practical aspects of deploying DCP. Section V presents the datasets used in our experiments and discusses the performance results, demonstrating the effectiveness of DCP in various scenarios. Finally, Section VI concludes the paper, summarizing the key findings and outlining potential directions for future research.

## II. RELATED WORK

Chiang & Lee scrutinize the effectiveness and legitimacy of synonym substitution analyses in NLP, providing valuable perspectives on their capabilities and limitations [4]. By questioning the conventional view of these attacks, the authors contribute to a deeper comprehension of NLP security vulnerabilities.

Asl et al. propose a framework called SSCAE that is capable of crafting sophisticated adversarial examples in natural language processing [5]. However, the complexity of incorporating diverse linguistic features of the scheme may pose challenges in scalability and efficiency.

Vitorino et al. evaluate the efficiency of adversarial evasion attacks against large language models [6]. However, the proposition may face limitations in replicating real-world scenarios and assessing the long-term effectiveness of proposed countermeasures.

Zhao et al. introduce a method for generating adversarial alterations at the word level utilizing the differential evolution algorithm [7]. However, potential weaknesses may arise in the method's ability to generate diverse and robust adversarial examples across different datasets and model architectures.

Li et al. explore query-limited adversarial attacks targeting graph neural networks (GNNs) [8]. However, potential weaknesses may arise in the generalization and robustness of the proposed attack method across different GNN architectures and datasets.

Hu et al. propose *FastTextDodger*, a decision-based adversarial attack tailored for black-box NLP models [9]. However, potential limitations may arise in scenarios where the attack's effectiveness depends heavily on the target model's decision boundary complexity, necessitating further investigation into its generalization.

Parry et al. discuss various intricacies of adversarial attacks specifically tailored for sequence-to-sequence relevance models [10]. However, addressing these vulnerabilities effectively requires a nuanced understanding of the underlying model dynamics and the development of tailored defense mechanisms.

Waghela et al. introduce a novel scheme, MWSAA that enhances traditional word saliency-based attacks by incorporating modifications to optimize adversarial perturbations [11]. The authors illustrate how well their method works in creating adversarial samples that deceive text classification models while maintaining semantic coherence. However, further exploration is needed to evaluate the method's performance across diverse datasets, as well as its scalability to real-world applications. The same authors also designed another enhanced adversarial attack, SASSP, integrating saliency, attention, and semantic similarity [12]. Empirical evaluations demonstrate SASSP's efficacy in generating adversarial samples with high semantic fidelity and superior attack success rates.

Despite advancements in adversarial text generation, existing methods still face challenges in producing examples that are both effective and imperceptible. Many approaches focus solely on maximizing the model's prediction error without considering contextual relevance or semantic consistency, resulting in nonsensical or linguistically unnatural outputs that limit practical utility. DCP addresses these shortcomings by dynamically adapting perturbations based on the context of the text and optimizing for both misclassification likelihood and text fluency.

## III. METHODOLOGY

The DCP scheme aims to generate sophisticated adversarial text examples that challenge the robustness of state-of-the-art NLP systems. This section outlines the methodology of DCP in detail, providing a comprehensive explanation of each step involved in the process. Fig 1 depicts a flow diagram to visualize the workflow of DCP.

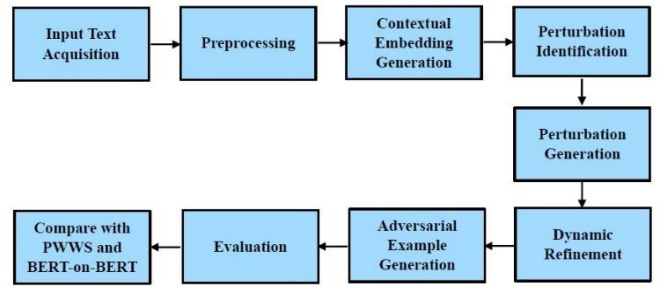

Fig. 1. The flow diagram of the steps involved in the design of DCP

*Step 1: Input Text Acquisition* - The initial step is acquiring the input text. For text classification tasks the datasets chosen for this study include IMDB and Yelp for sentiment analysis, AG News for topic classification, and Fake News for classifying news articles as fake or genuine. For natural language inference tasks, MNLI and SNLI datasets are used. These datasets are publicly available.

*Step 2: Preprocessing* - Once the input text is collected, it undergoes preprocessing. Tokenization splits the text into individual tokens, making it easier to analyze. Lowercasing ensures uniformity, while stop-word removal eliminates common words that do not significantly affect the meaning. Stemming or lemmatization reduces words to their root forms, helping to standardize different variations of the same word. Noise removal clears out any non-alphabetic elements, ensuring a clean text input for subsequent steps.

*Step 3: Contextual embedding generation* - Pre-trained language models like BERT are used to generate contextual embeddings for the preprocessed text. These embeddings capture the semantic and syntactic nuances of words within their specific contexts.

*Step 4: Perturbation identification* - Identifying perturbation candidates involves calculating the gradient of the loss function concerning the input text. This helps pinpoint words that have the most influence on the model's output. Saliency maps are used to identify keywords and phrases that contribute to the classification task.

*Step 5: Perturbation generation* - Generating context-aware perturbations is the next step. Synonym substitution replaces words with contextually appropriate synonyms using resources like WordNet or predictions from masked language models.

*Step 6: Dynamic refinement* - Dynamic refinement is an iterative process that balances the objectives of causing misclassification and maintaining readability. The misclassification objective maximizes the loss of the model $L_{model}(x + \delta, y; \theta)$, where $\delta$ is the perturbation applied to the input text $x$. This increases the likelihood of misclassification. To maintain readability, the function minimizes the difference between the original and perturbed text embeddings using $L_{sim} = \|E(x) - E(x + \delta)\|_2^2$ ensuring semantic similarity. The combined objective function is given by (1)

$$L_{adv} = L_{model}(x + \delta, y; \theta) + \lambda * L_{sim} \quad (1)$$

In (1), $\lambda$ is the parameter that controls the trade-off between misclassification and readability. A higher value of $\lambda$ prioritizes readability, while a lower value favors misclassification. This optimization technique is used to iteratively refine the perturbations.

*Step 7: Adversarial example generation* - After refining the perturbations, the final adversarial examples are generated. These examples are validated to ensure they meet the criteria for misclassification and semantic consistency.

*Step 8: Evaluation* - The effectiveness of the generated adversarial examples is evaluated using metrics such as attack success rate (ASR), perturbation magnitude, and readability. The ASR measures the percentage of adversarial examples that successfully cause misclassification. Perturbation magnitude evaluates the extent of changes made to the original text, while *readability* assesses the readability of the adversarial examples using the Flesch-Kincaid score.

*Step 9: Comparison with PWWS and BERT-on-BERT* - To compare the performance of DCP with PWWS [13] and BERT-on-BERT [14] attacks, the same datasets and models are used. This involves implementing PWWS and BERT-on-BERT attacks on the preprocessed datasets and generating adversarial examples.

## IV. IMPLEMENTATION

This section provides the details of the implementation of the DCP scheme in the Python programming language.

Fig 2 exhibits the pseudocode for the algorithm of the DCP scheme.

*Step 1: Setup and preprocessing* - This step involves the following tasks: (a) installing libraries, (b) initializing NLTK, and (c) designing the *setup and preprocessing* function. The installed libraries include *nltk*, *transformers*, *datasets*, and *torch*. The *nltk* library provides resources for NLP tasks such as tokenization, stemming, tagging, and parsing. The *transformers* library provided by Hugging Face offers an easy-to-use interface for working with pre-trained transformer models like BERT, GPT, and others. The *datasets* library of Hugging Face facilitates easy access to various datasets for NLP tasks. The *torch* library is used for building and training neural networks, particularly in deep learning applications.

```
# Step 1: Input Text Acquisition
input_text = get_input_text(dataset)
# Step 2: Preprocessing
preprocessed_text = preprocess(input_text)
# Step 3: Contextual Embedding Generation
embeddings = generate_contextual_embeddings(preprocessed_text, pre_trained_model="BERT")
# Step 4: Perturbation Identification
# Compute gradients and identify important words for perturbation
important_words = identify_perturbation_candidates(preprocessed_text, model, embeddings)
# Step 5: Perturbation Generation
# Generate context-aware perturbations using synonym substitution or paraphrasing
perturbed_texts = generate_perturbations(important_words, preprocessed_text, method="synonym_substitution")
# Step 6: Dynamic Refinement
# Initialize adversarial objective and iteratively refine perturbations
while not misclassification(perturbed_texts, model):
    for each perturbed_text in perturbed_texts:
        # Compute misclassification loss
        loss_misclassification = compute_loss(perturbed_text, model)
        # Compute similarity between original and perturbed text (readability)
        loss_similarity = compute_similarity(preprocessed_text, perturbed_text)
        # Update perturbation based on adversarial objective
        adversarial_objective = loss_misclassification + lambda * loss_similarity
        perturbed_text = refine_perturbation(perturbed_text, adversarial_objective)
# Step 7: Adversarial Example Generation
# Validate final adversarial examples
adversarial_examples = validate_adversarial_examples(perturbed_texts)
# Step 8: Evaluation
evaluate(adversarial_examples, original_text=input_text, metrics=["ASR", "perturbation_rate", "readability"])
```

Fig. 2. The pseudocode for the steps involved the DCP algorithm

*Step 2: Generate Contextual Embeddings* – The contextual embeddings for the input text are generated using a pre-trained BERT model. Before generating embeddings, the pre-trained BERT model and *tokenizer* from the Hugging Face's Transformers library are loaded. The tokenizer is used to convert text into a format that the BERT model can understand, typically by splitting the text into tokens and mapping them to their corresponding token IDs. The pre-trained BERT model is loaded using *model = BertModel.from_pretrained('bert-base-uncased')*. The BERT model will generate embeddings for the input tokens. The function *get_embeddings* takes a text input, tokenizes it, and generates contextual embeddings.

*Step 3: Perturbation identification* – This involves identifying which words in the text should be perturbed based on their gradients concerning the model's output. The function *calculate_gradient* performs this task as follows. The input text is tokenized and prepared for the model. Gradient computation is enabled for the input tokens. A forward pass is performed through the model to compute the outputs. The loss is computed, and backpropagation is performed to calculate the gradients. The absolute values of the gradients are summed across the embedding dimensions to get a single importance score for each token. Finally, the tokens with the highest gradient magnitudes are identified.

*Step 4: Perturbation generation* – The first step in perturbation generation involves finding synonyms for the words that have been identified for perturbation. The function *get_synonym* performs the task by executing *task.wordnet.synsets(word)*, and retrieving all *synsets* (sets of synonyms) for the given word. For each *synset*, the function *get_synonym* iterates through the *lemmas* (individual word forms) and adds them to the *synonyms* set. Once the synonyms are available, the perturbed versions of the input text are generated by perturbing the identified important words with their synonyms by the function *generate_perturbations*.

*Step 5: Dynamic refinement* - It iteratively generates perturbed text until the adversarial objective is achieved. The *adversarial_objective* function evaluates whether the perturbed text achieves the adversarial objective, i.e., misclassification. It uses a pre-trained classifier based on BERT to classify the perturbed text and compares the predicted label with the original label. The *dynamic_refinement* function iterates over a loop where it repeatedly generates perturbed text until the adversarial objective is achieved.

*Step 6: Evaluation* - This computes various metrics to evaluate the success of the perturbations and their impact on the original text. The evaluation metrics include (a) attack success, (b) perturbation rate, and (c) semantic similarity. Attack success indicates whether the attack successfully caused misclassification. The perturbation rate measures the percentage of words changed in the original text.

## V. Performance Results

The effectiveness of the DCP attack scheme is evaluated and contrasted with the PWWS attack [13] and the BERT-on-BERT attack [14]. In line with BERT-Attack [14], we evaluate PGD-BERT using 1000 test examples randomly chosen from the respective task's test data set, consistent with partitions used in [3]. The comparative study with PWWS includes the following pre-trained models: (i) Word-CNN [15], (ii) Bi-LSTM [16], and (iii) Char-CNN [17]. The datasets used in this study include AG News [18], IMDB [19], Yelp [20], Fake News [21], MNLI [22], and SNLI [23]. For comparing the performance of DCP with BERT-on-BERT, Word-LSTM [24], BERT-Large [25], and ESIM [26] are also used.

It is observed from Tables I and II and Fig 3 that CNN and LSTM models generally maintain high accuracy in the absence of attacks for AG News and IMDB datasets on the fine-tuned BERT victim model. However, the accuracy significantly drops in the presence of attacks. The adverse effect of DCP is even more as it leads to lower accuracy. Moreover, DCP results in lower perturbation rates compared to PWWS, implying it can generate more semantically similar adversarial examples with fewer modifications.

TABLE I. Classification Accuracy in Absence of any Attack and Presence of PWWS and DCP Attacks for AG News and IMDB

| Dataset | Model | Acc in abs of Attack (%) | Acc. under PWWS (%) | Acc. under DCP (%) |
|---|---|---|---|---|
| AG News | Word-CNN | 90.56 | 56.72 | 48.25 |
| | Char-CNN | 89.70 | 56.20 | 46.20 |
| IMDB | Bi-LSTM | 84.86 | 2.20 | 1.75 |
| | Word-CNN | 86.55 | 5.50 | 3.60 |

TABLE II. Perturbation Rate of PWWS and DCP Attacks for AG News and IMDB

| Dataset | Model | Perturb Rate with PWWS (%) | Perturb Rate with DCP (%) |
|---|---|---|---|
| AG News | Word-CNN | 16.76 | 15.25 |
| | Char-CNN | 18.93 | 14.80 |
| IMDB | Bi-LSTM | 3.38 | 2.80 |
| | Word-CNN | 3.81 | 3.10 |

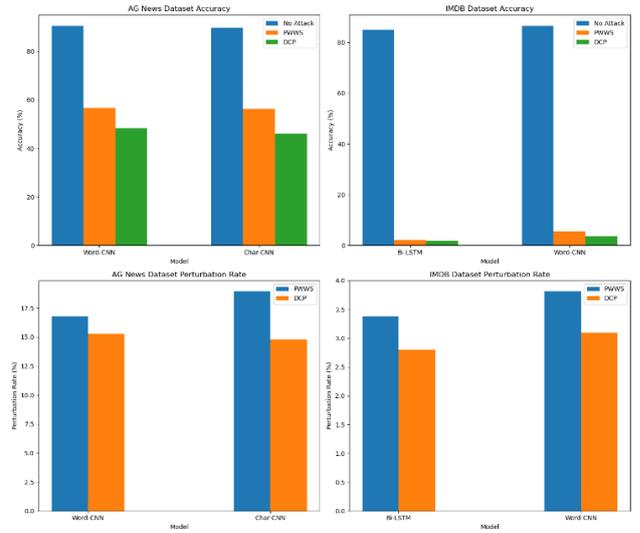

Fig. 3. The classification accuracies and perturbation rates for PWWS and DCP attacks for different models on AG News and IMDB datasets

TABLE III. Accuracy and Perturbation Rate for BERT-on-BERT and DCP Attacks on Different Datasets for Text Classification

| Dataset | Attack Method | Original Accuracy | Accuracy in Presence of Attack | Perturb Rate of the Attack |
|---|---|---|---|---|
| IMDB | BERT-on-BERT | 90.90 | 11.40 | 4.40 |
| | DCP | | 7.40 | 2.70 |
| Yelp | BERT-on-BERT | 95.60 | 5.10 | 4.10 |
| | DCP | | 4.05 | 3.50 |
| Fake | BERT-on-BERT | 97.80 | 15.50 | 1.10 |
| | DCP | | 11.4 | 0.90 |
| AG News | BERT-on-BERT | 94.20 | 10.60 | 15.40 |
| | DCP | | 6.70 | 8.60 |

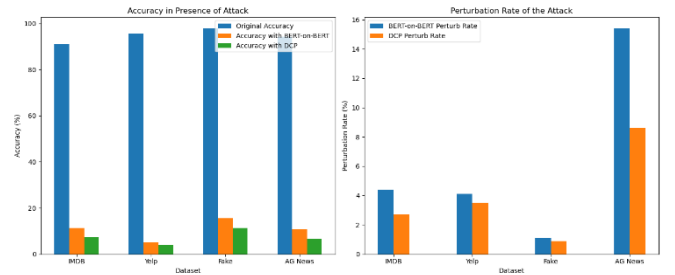

Fig. 4. Classification accuracy and perturbation rate for BERT-on-BERT attack and DCP attack for various datasets for text classification

Table III and Fig 4 depict the accuracies and perturbation rates of BERT-on-BERT and DCP attacks on a fine-tuned BERT model for several datasets for text classification tasks. While BERT-on-BERT and DCP both significantly lower the model's accuracy across all datasets, the effect of DCP is more severe. Moreover, DCP achieves this with consistently lower perturbation rates, making the attack more subtle and potentially more dangerous.

TABLE IV. NUMBER OF QUERIES GENERATED AND SEMANTIC SIMILARITY FOR BERT-ON-BERT AND DCP ATTACKS ON TEXT CLASSIFICATION DATA

| Dataset | Attack Method | No of Queries | Semantic Similarity |
|---|---|---|---|
| IMDB | BERT-on-BERT | 454 | 0.86 |
|  | DCP | 347 | 0.96 |
| Yelp | BERT-on-BERT | 273 | 0.77 |
|  | DCP | 238 | 0.94 |
| Fake | BERT-on-BERT | 1558 | 0.81 |
|  | DCP | 943 | 0.93 |
| AG News | BERT-on-BERT | 213 | 0.63 |
|  | DCP | 154 | 0.94 |

Table IV and Fig 5 show that DCP requires fewer queries to generate adversarial examples while consistently maintaining higher semantic similarity in the perturbed texts compared to BERT-on-BERT for text classification tasks.

Table V and Fig 6 present the performance results of BERT-on-BERT and DCP attacks on several datasets for natural language inference tasks. The attack results for hypotheses (H) and premises (P) are shown separately. DCP is found to consistently outperform BERT-on-BERT.

Table VI depicts the impact of BERT-on-BERT and DCP attacks on different models other than fine-tuned BERT models. The results demonstrate the effectiveness of these attacks across different datasets and model architectures. However, DCP demonstrates a superior capability to compromise the models in comparison to BERT-on-BERT.

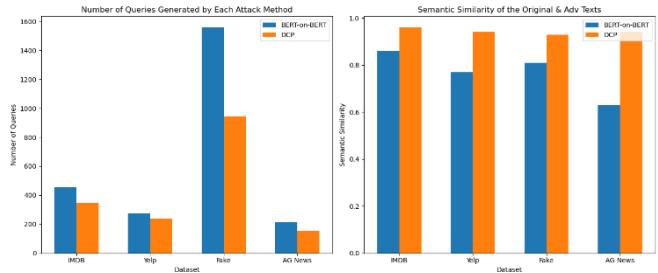

Fig. 5. Number of queries generated and semantic similarity in the texts for BERT-on-BERT and DCP attacks on text classification datasets

TABLE V. ACCURACY AND PERTURBATION RATE FOR BERT-ON-BERT AND DCP ATTACKS ON NATURAL LANGUAGE INFERENCE DATASETS

| Dataset | Attack Method | Original Accuracy | Accuracy in Presence of Attack (H/P) | Perturb Rate of the Attack (H/P) |
|---|---|---|---|---|
| MNLI Matched | BERT-on-BERT | 85.10 | 7.90/11.90 | 8.80/7.90 |
|  | DCP |  | 5.30/10.80 | 7.40/6.70 |
| MNLI Unmatched | BERT-on-BERT | 82.10 | 7.00/13.70 | 8.00/7.10 |
|  | DCP |  | 5.10/10.60 | 7.20/7.00 |
| SNLI | BERT-on-BERT | 89.40 | 7.40/16.10 | 12.40/9.30 |
|  | DCP |  | 3.20/12.60 | 8.20/6.30 |

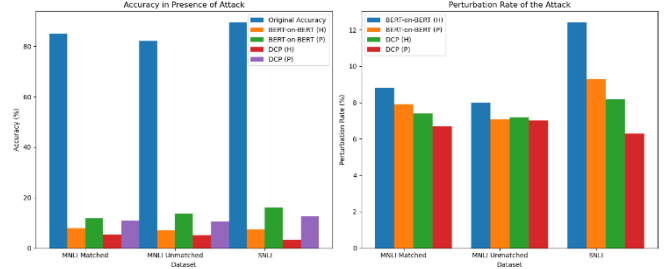

Fig. 6. Accuracy and perturbation rate for BERT-on-BERT and DCP attacks on natural language inference-related datasets

TABLE VI. ATTACK TRANSFERABILITY PERFORMANCE – ACCURACY OF BERT-ON-BERT AND DCP ATTACKS ON DIFFERENT MODELS

| Dataset | Model | Acc in Absence of Attack | Acc in Presence of BERT-on-BERT | Acc in Presence of DCP |
|---|---|---|---|---|
| IMDB | Word-LSTM | 89.80 | 10.20 | 7.40 |
|  | BERT-Large | 98.20 | 12.40 | 8.30 |
| Yelp | Word-LSTM | 96.00 | 1.10 | 0.70 |
|  | BERT-Large | 97.90 | 8.20 | 5.40 |
| MNLI Matched | ESIM | 76.20 | 9.60 | 7.20 |
|  | BERT-Large | 86.40 | 13.20 | 10.80 |

## VI. CONCLUSION

This paper introduced DCP, a novel adversarial attack scheme for NLP models. DCP generates context-aware perturbations that maintain semantic fidelity and fluency, leveraging pre-trained language models. Experimental results show DCP's effectiveness across multiple NLP models and datasets, achieving higher attack success rates with minimal perturbations than other methods. This study highlights the importance of context in adversarial text generation. Future work will enhance defense mechanisms, explore broader NLP applications, and address the ethical implications of adversarial attacks.